# The Role of Temperature on Irradiation Defect Evolution near Surfaces and Grain Boundaries in Tungsten

Yang Zhang[1] and Jason R. Trelewicz[1,2,*]

[1]Department of Materials Science and Chemical Engineering, Stony Brook University, Stony Brook, NY 11794

[2]Institute for Advanced Computational Science, Stony Brook University, Stony Brook, NY 11794

*Corresponding Author at Stony Brook University: jason.trelewicz@stonybrook.edu

## Abstract

This study explores the impact of temperature on defect dynamics in tungsten, emphasizing its application in nuclear fusion reactors as Plasma Facing Components (PFCs). Through atomistic simulations, the research elucidates the intricate interplay of defect production, annihilation, and redistribution under irradiation at room (300K) and elevated temperatures (1000K). It demonstrates that higher temperatures significantly increase the number and mobility of defects, leading to a substantial rise in the total number of surviving Frenkel Pairs (FPs), with a notable preference for surface distribution. This redistribution is attributed to energy gradient-driven relocation processes, enhanced by the defects' increased mobility at elevated temperatures. Moreover, the study reveals that elevated temperatures promote biased accumulation of interstitial defects in grain boundaries, especially in configurations that facilitate efficient interstitial migration, indicating a strategy for minimizing lattice interstitial accumulation under irradiation. These findings underscore the critical role of temperature in modulating irradiation-induced defect dynamics in tungsten, providing valuable insights for designing and selecting materials with optimized irradiation resistance for use in extreme conditions of nuclear fusion reactors. The research suggests a material design approach that accounts for temperature effects to enhance the durability and performance of nuclear fusion materials.

Tungsten is one of the most important candidate materials for Plasma Facing Components (PFCs) in future nuclear fusion reactors[1-4], due to its high thermal conductivity, sufficient life time due to erosion, negligible/low outgassing, acceptable neutron resistance[3-12]. In ITER-like devices, those PFCs will be working under the extreme conditions, such as 1000K or higher temperature, and up to 50 dpa/year neutron irradiation. Seeking for the practical regulations of the material implementation under these conditions, the damage production and accumulation mechanisms need to be described as detailed as possible.

In this nanometer, nanosecond field, the atomistic simulation provides an irreplaceable tool. A number of studies have been completed to understand the defect evolution during isolated impact event[13-17], these findings along with the research on grain boundary effects on defect evolution[18-28] enlightened the studies of nanocrystalline (NC) material as one with potentially better irradiation resistance. Inspired by those previous works, Zhang et al took one step further, a study combined the surface effects, the grain boundary effects and the accumulation effects provided a detailed story of the material under cumulative irradiation at the room temperature, their results suggested that the irradiation damage accumulation in the material was a complicated process including the thermodynamic vacancy production process, the kinetic interstitial production process, and the energy gradient driven interstitial migration process.

While those findings revealed some fundamental mechanisms of the defect evolution, it would be necessary to compare the results at the room temperature and at the typical working temperature before giving any practical predictions, as previous works had shown behavior shifts at different temperatures[16, 28, 29]. Based on the same irradiation simulation methods, this work provided a direct comparison between the results at 1000K and the results at 300K. The temperature depicted profound, and biased effects on defect production, annihilation, and redistribution behaviors.

As shown in Figure 1a, a single 1KeV self-atom impact event can be divided into two parts: the collision cascade developing period and the defect recombination period. The kinetic energy of the PKA passed to the atoms in a small region upon impact, causing a collision cascade, a peak damage region with several Replacement Collision Sequences (RCSs) spreading out. Around 1ps after the impact, the collision cascade reached its peak with several dozens of Frenkel Pairs (PFs), after that the interstitials and the vacancies started to recombine with each other, causing the reduction in total number of FPs.

As a thermodynamic process, higher total energy would result in a higher total number of FPs produced. The background temperature determined the base energy of the atoms, and reduced the formation energy of interstitials and vacancies. The collision cascade evolutions at two different temperatures showed the consistency with the prediction. Higher number of peak FPs were observed at the elevated temperature, shown in Figure 1b. And the recombination process depended on the redistribution of defects, with higher mobility at higher temperature, the recombination efficiency was also observed to be higher at the elevated temperature.

Combining these two competing effects together, the results showed a significant increase (50%~60%) of the total surviving FPs, including the lattice defects as well as the surface defects. However, a closer look revealed more profound details, the defects was mainly distributed on the surface. The comparison between the peak state and the final state of the collision cascade shown in Figure1c and d, this kind of defect distribution should be a result of the defect relocation process, thus the direction of the relocation should be energy gradient dependent, and the efficiency of the relocation should be mobility related.

The difference of the formation energy between the lattice interstitial and the surface interstitial was significant, 6.2eV and 5.7eV for (110) surface at 300K and 1000K respectively, and 9.1eV and 8.4eV for (100) surface at the different temperatures. The energy gradient was smaller at the elevated temperature while the fraction of surface interstitial was increased. This indicated the relocation efficiency was more affected by the mobility. As a result, the fraction of surface vacancy was smaller than the fraction of surface interstitial, or the fraction of lattice vacancy was higher than the fraction of lattice interstitial, and the temperature intensified the relocation.

Thus, the temperature effect on the defect evolution by 1KeV impact on a surface can be summarized as: (i) the peak number of defects, the recombination would increase with the temperature, (ii) the total number of surviving defects increased significantly when the temperature increased from 300K to 1000K, (iii) because of the energy gradient, most of the defect relocated to the surface after production, (iv) the mobility controlled relocation process determined that the relocation efficiencies of vacancy and interstitial were different, and the elevated temperature had positive effect on the relocation process.

As talked in the study on the cumulative irradiation at room temperature, the overlap effect would divide the simulation into two parts: the damage proliferation period, and the damage saturation period. During the first 50 impacts, the impacts seldom hit the damaged area, and the number of defects accumulated fast. When the surface of the target was fully damaged, the successive impacts created significantly smaller number of new defects, and the defect accumulation decelerated to almost zero in the damage saturation period. It was worth noting here the close to zero defect accumulation rate didn't mean that the impacts stopped producing defects, but the defect annihilation rate was close to the production rate.

The discussion started from the simpler version, in the 12nm grain size structures, the cascades would not interact with the GBs much. In these structures, basic defect behaviors were consistent when the cumulative irradiation was performed at the elevated temperature (Figure 2). While the vacancy accumulation rate was reduced at 1000K, the interstitial accumulation rate was increased. The slower vacancy accumulation rate would relate to the basic nature of the cascade evolution at the elevated temperature talked above (Figure 1b). And the increased interstitial accumulation rate should be a result of the reduced interstitial annihilation rate (Figure 2). The reduced cascade size and the enhanced mobility helped the interstitials to move deeper into the material and farther away from the vacancy-rich region, thus decreased the frequency of interstitial annihilation, which majorly caused by the FP recombination.

The snapshots taken during the simulation (Figure 2c&d) showed that the dominant FP recombination processes differed when the background temperature changed from 300K to 1000K. At 300K, the interstitials needed to be energized by the impact shockwave, which required the cascade to develop near the interstitials. Figure 2c showed the atomic displacements happened during one impact event: some of the pre-existing vacancies were filled by the atoms from the cascade core, and some of the pre-existing interstitials displaced towards the cascade core. At 1000K, the interstitials were mobile without activation. Figure 2d was a combined image of 5 snapshots, the time between subsequent snapshot was 16ps, summing up to 80ps. The constant interstitial migrations were observed in the region away from the cascade region.

To have a whole image of the migration behavior, the snapshots were taken at the end of the simulation (Figure 3). The picture of the atomic displacements supported the idea that the difference of the interstitial distribution at different temperatures would likely to be a result of the different interstitial mobility, since at the elevated temperature (i) extensively more displacements were observed in the structure, (ii) the interstitials were distributed farther away from the bombardment area, and (iii) none of the interstitial remained adjacent to the vacancy-rich region. Thus, at the elevated temperature, though the recombination was enhanced during the cascade evolution, the recombination happened at a lower rate after the active phase.

The snapshots (Figure 3) also supported the illustration on the lattice vacancy behavior. As the previous work implicated, the behavior of the lattice vacancy accumulation can be estimated by the size of the total "interaction volume". Here, the "interaction volume" was defined as the region where the atoms displaced more than 2.8Å averaging over the first nearest neighbor, where the 2.8Å was determined by the length between nearest neighbors. Based on this definition, in the interaction volume, the majority of the atoms were displaced from their original sites at the end of the simulation. Consistent with what Figure 1 indicated, the interaction volume of the self-atom irradiation was more localized at the elevated temperature. which was one cause of the reduced lattice vacancy accumulation rate.

When the GB effects were included in the 6nm grain size structures (Figure 4a&b), the cascade-GB interactions promoted the interstitial accumulation in the GB, and the interstitial accumulation rate was even higher in the GB than in the lattice. This biased interstitial accumulation created an extremely high interstitial concentration difference between the GB and the lattice. Despite the concentration difference, the interstitial annihilation contributed by the GB interstitials was neglectable compared with the lattice interstitials, which was a consequence of two causes, (i) the GB interstitials were farther away from the vacancy-rich region, (ii) the migration barrier from the GB to the lattice was high.

Additionally, the temperature had different impacts on the mobility of interstitials in different types of GBs (Figure 4c). At the end of the simulations in two Σ3-6nm structures, 5 times more atoms were displaced at 1000K than at 300K. These displacements recorded the motion of crowdions, as the snapshots of the GB planes (Figure 4c) revealed. In other words, the displacements snapshots showed that while the interstitials in the Σ3 GB were significantly more mobile at 1000K than at 300K, the mobility of the interstitials in the Σ5 GB was not affected much by the temperature. As a result, it was much easier for the interstitials in the Σ3 GB to redistribute at 1000K, and this redistribution ability provided an explanation for the promoted GB interstitial production in the Σ3 GB at 1000K (Figure 4d): The redistribution promoted the interstitial clustering in the GB (Figure 5a), which reduced the concentration of the interstitials elsewhere, as a result, the successive interstitials had a higher probability to migrate into the GB without impediment from pre-existing interstitials.

The elevated temperature not only promoted the interstitial clustering in the GB, but also resulted in the interstitial clustering in the lattice. Without the constraint from the GB, an interstitial dislocation loop was observed in the Σ5-12nm structure during the irradiation at the elevated temperature (Figure 5b&c), the interstitial cluster maintained the mobility as other studies predicted[30]. The simulation snapshots in the present study showed that the velocity of the interstitial dislocation loop during migration can be as high as 218m/s at the elevated temperature.

As a consequence of the all the mechanisms described above, the temperature effect on the defect evolution during low energy self-atom irradiation can be summarized (Figure 6): (i) the saturation numbers of vacancy decreased at the elevated temperature in all cases we tested, because of the enhanced surface-defect interaction during the cascade evolution and reduced cascade size. (ii) the number of lattice interstitials decreased as the temperature increased, which was a result of the promoted interstitial redistribution at the elevated temperature, as well as the interstitial clustering. (iii) the elevated temperature generally increased the fraction of GB interstitials, however the temperature effects differed in different GB configurations: the elevated temperature increased the number of GB interstitials in the $\Sigma 3$ GB, which had lower interstitial migration barrier, and in which the interstitial clustering was observed.

These detailed atomistic simulation observations added to the understanding of the irradiation defect evolution under the working condition, and provided relevant information in choosing a better configuration of the irradiation resistant material: the simulation results indicated that a structure with smaller grain size, and GBs allowing the interstitial to migrate inside, accumulated the least lattice interstitials during the low energy self-atom irradiation.

**Acknowledgements**

**Figures**

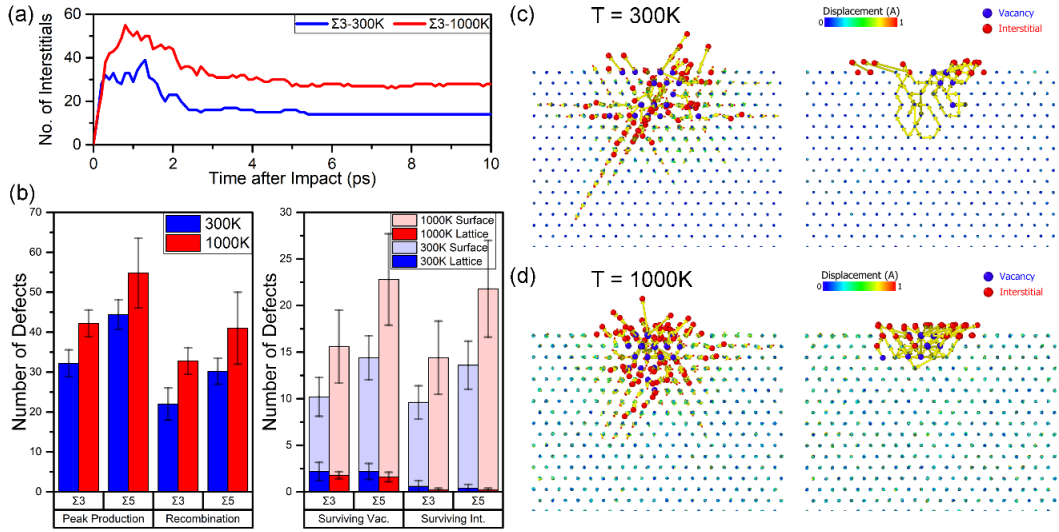

Figure 1 The average defect production by 1KeV impact on the structure with surface at different temperatures. The biased distribution of vacancy and interstitial. The fraction of vacancies created as surface vacancies differs from the fraction of interstitials created as surface interstitials, and at the elevated temperature the fraction of surface defects increased.

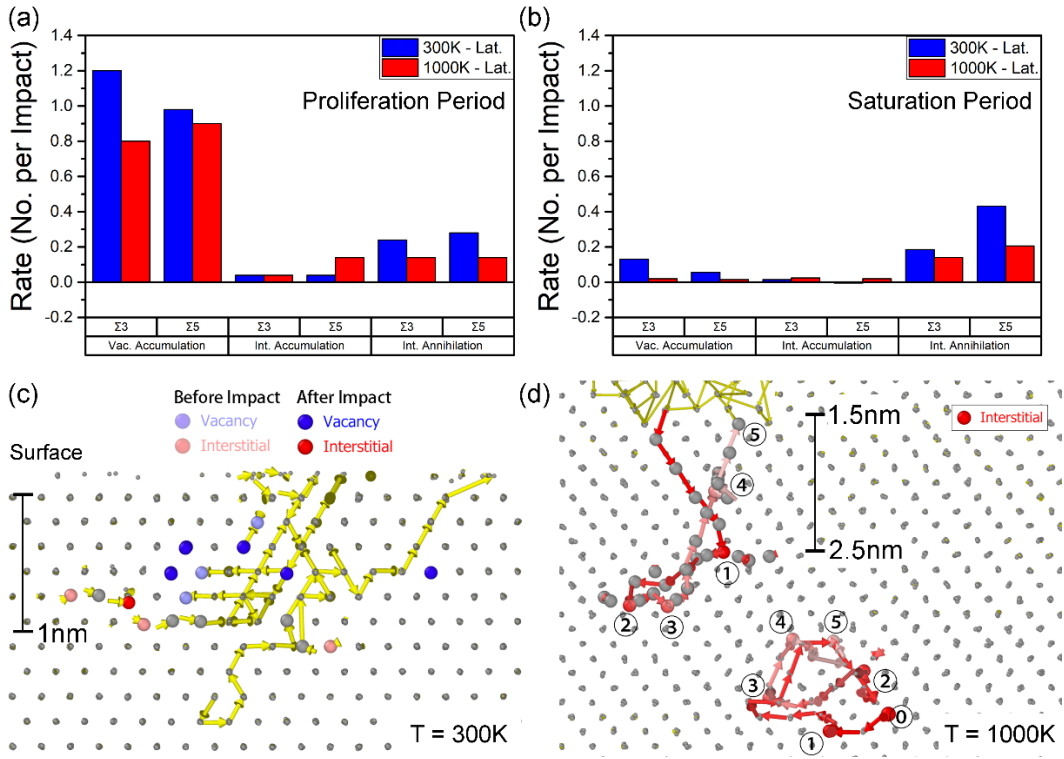

Figure 2 **upper figures**: The defect accumulation and annihilation rate in 12nm grain size structures during the defect proliferation period and defect saturation period. The vacancy accumulation rate is higher at the room temperature, while the interstitial accumulation rate is higher at the elevated temperature. **lower figures**: a) Most common process for the interstitial annihilation happened when a successive collision cascade developed near interstitial. The CC provided kinetic energy to the interstitial nearby, enabled them to cross the energy barrier and recombine with the nearby vacancies. The interstitial annihilation rate was also higher during the damage saturation period, since the overlap probability was higher, and the CCs would have a higher chance to develop near interstitials. b) Low interstitial annihilation rate at high temperature related to the higher mobility of the interstitial. The interstitials would distribution farther away from the Peak Damage Region of the Collision Cascade. Even if the impact hit on the previous damaged area, the radius of the CC may not be enough to activate the interstitials.

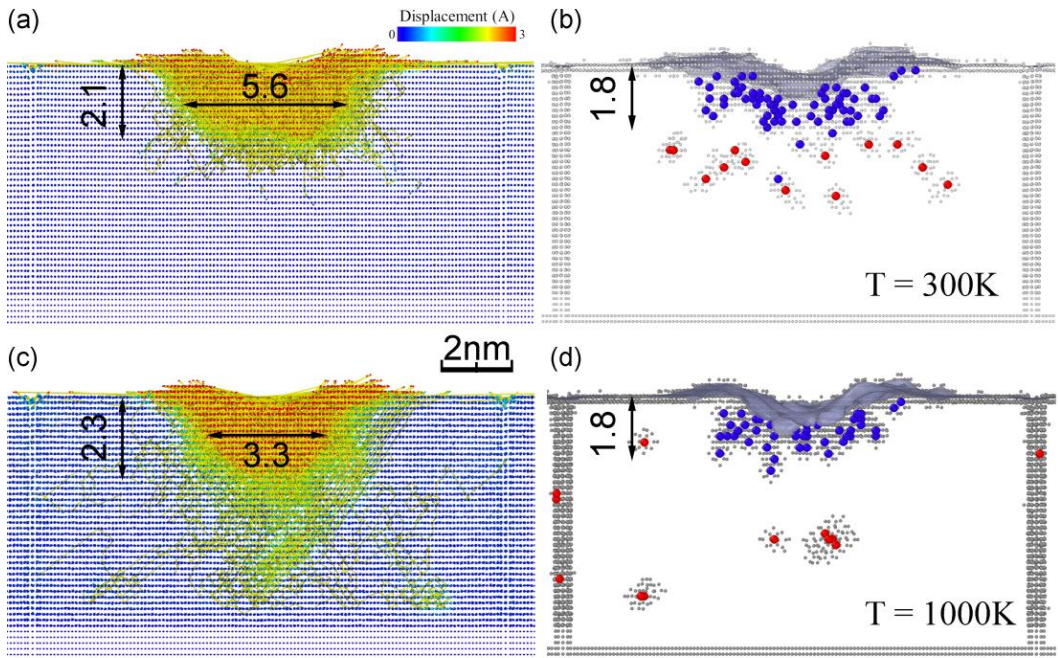

Figure 3 Defect distribution at the end of the cumulative irradiation at different temperatures.

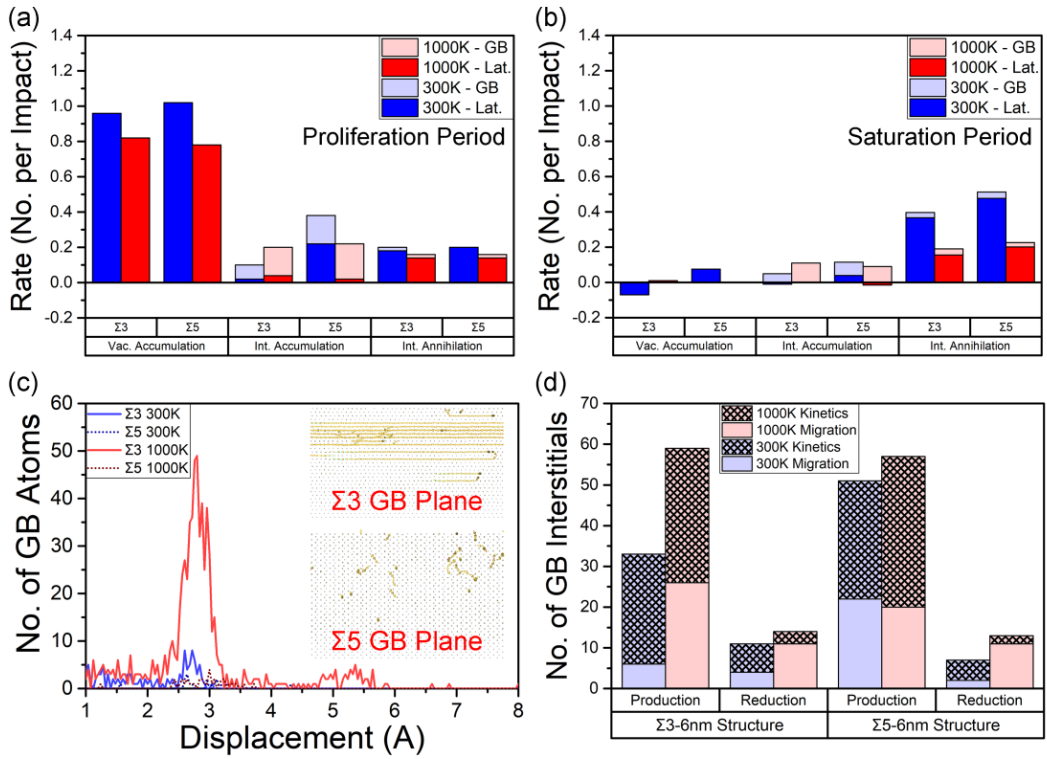

Figure 4 The defect accumulation and annihilation rate in 6nm grain size structures during the defect proliferation period and saturation period. The interstitial annihilation process took place more frequently in the lattice than in the GB, while the interstitial accumulation in the GB was as significant, or even higher than the interstitial accumulation in the lattice. The GB interstitial accumulation was exacerbated at the elevated temperature, while the temperature had trivial influence on the GB interstitial annihilation. (c) The interstitial in the Σ3 GB was much more mobile than the interstitial in the Σ5 GB. Due to the high mobility, the interstitials in the Σ3 GB would be redistributed and structured a lower energy configuration to keep these interstitials. Because of this temperature dependent redistribution, the interstitial behaved more different at different temperatures near Σ3 GB than near Σ5 GB, a higher increase of the production of GB interstitials from migration process was observed in the Σ3 GB. For the Σ5 GB, the result indicated that the temperature affected less significantly on GB interstitial production, probably the wide and deep energy well near the Σ5 GB already applied significant amount of driving force on the interstitials and the interstitial capacity of the GB was limited.

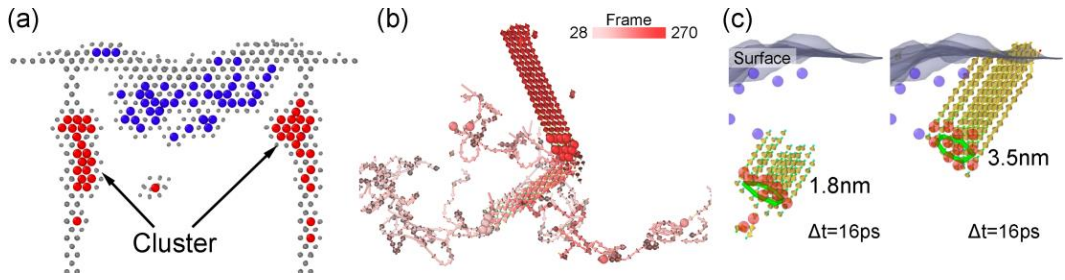

Figure 5 At the elevated temperature, (a) the GB interstitial clusters in Σ3-6nm structure at the end of the simulation. (b) the migration induced interstitial clustering. (c) dislocation loop analysis revealed the interstitial cluster (b) in the form of dislocation loop.

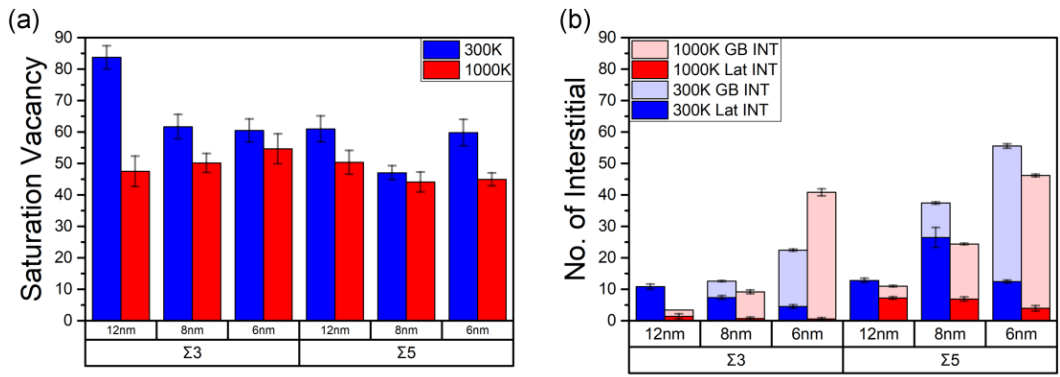

Figure 6 (a) Number of Vacancy averaged over the last 50 impacts. (b) Number of Interstitials in Lattice and Grain Boundary averaged over the last 50 impacts.

The elevated temperature lowered the saturation number of vacancies. And it reduced the fraction of the interstitials in lattice more significantly.

The temperature had profound effects on the defect production, annihilation and distribution processes. The surface, grain boundary as well as the vacancies were competing each other in collecting interstitials, and the competition was exacerbated at the elevated temperature.